\documentclass{scrartcl}
\usepackage[affil-it,auth-lg]{authblk}

\usepackage[latin1]{inputenc}
\usepackage{amsmath}
\usepackage{amsfonts}
\usepackage{amssymb}
\usepackage{bm,amsmath,amssymb,amsfonts,latexsym,psfrag,pstricks}
\usepackage{bm}
\usepackage{amssymb}
 \usepackage{array}
\usepackage{mathtools}
\usepackage{amsthm}
\usepackage{newlfont}
\usepackage{esdiff}
\usepackage{multirow}

\usepackage{graphicx}
\usepackage{parskip}
\usepackage{graphicx}
\usepackage{url}                
\usepackage{bm,amsmath,amssymb,amsfonts,latexsym,psfrag}
\usepackage[small,it]{caption}
 \usepackage{amssymb}
 \usepackage{enumerate}
 \usepackage{color}
 \let\origthanks\thanks\renewcommand\thanks[1]{\begingroup\let\rlap\relax\origthanks{#1}\endgroup}

\title{Goethe's Faust and the Minimal Action Principle: a possible Universal  message } 
\author[a]{Claudio Verzegnassi\footnote{E-mail: claudio@ts.infn.it}}
\author[c]{Euro Spallucci\footnote{E-mail: \texttt{spallucci@ts.infn.it}}}
\affil[a]{Dipartimento di Chimica, Fisica e Ambiente, Universit\`a  di Udine, Italy}
\affil[c]{Dipartimento di Fisica, Sezione Teorica, Universit\`a degli Studi di Trieste and INFN, 
Sezione di Trieste, Strada Costiera 11, I-34151 Trieste, Italy}

\date{\large\today}

\begin{document}

\maketitle

\begin{abstract}
This short note is devoted to non-proficient in physics. Its purpose is that of proposing a possible universal connection between
the definition of Action given by Goethe in his ``~\emph{Faust}~'' tragic play and the  Hamilton's Principle of the \emph{Minimal Action}.  
This Principle is one of the most general ways of formulating the dynamics of a physical system both in classical and in 
quantum physics. The recent discovery of the long time searched Higgs particle may be seen as the most spectacular successful 
confirmation of the Standard Model formalism based on the Minimal Action Principle. 
Supported  by this  experimental confirmation, 
we  propose a possible general  meaning of the Principle, inspired by the Goethe's ``~\emph{Faust}~'' tragic play.
\end{abstract}

\section{Introduction}
One of the several reasons why Wolfgang Goethe is still remembered by a number of people after such a long time is probably his proposed 
solution to the conflicting visions of the original structure of the Universe: \emph{Chaos},
as seen by  Hesiod \cite{theogony} or \emph{Logos} (see for instance \cite{logos}), as seen by Saint John. According to Goethe, none of the two views was correct. 
His  proposal was that of a quantity that he named \emph{Action}, that should have acted as a mediator of the two opposite 
extremes that exist in any human being, in particular in the main character of his most famous tragic play, Doctor Faust.
\cite{faust}. \\
The actual details of this ``~mediation~'' are in our opinion  fascinating, and will be briefly summarized 
in the final part of this note, where we shall compare them with the analogous situation that can be observed  
nowadays in Physics. The reason of this comparison is that, 
in the historical developments of the Modern (Relativistic Quantum) Physics, there appears a quantity called ``~Action~''
 of a system which has a fundamental relevance, 
determining its possible time evolution in the presence of any of the existing known Forces via the Hamilton's Principle\cite{hamilton} . 

\section{Hamilton's Principle}
The starting point of our investigation is the  fundamental Hamilton's Principle :
\begin{center}
\emph{ Principle of Minimal  Action}.
\end{center}
For a general material system, introducing a Lagrangian function defined as
\begin{equation}
\mathcal{L}= T-V
\end{equation}
where $T$ is the kinetic energy and $V$ the potential one, the Principle states that, when the system evolves from an initial time 
$t_1$ to a final time $t_2$ along the ``~correct~'' path fixed by the present forces, the integral of the Lagrangian from $t_1$ 
to $t_2$, i.e., in a less mathematical language, the ``~sum~'' of the Lagrangian values at all times in the 
interval $(t_1\ , t_2)$, is ``~\emph{minimum}~''
\footnote{ More generally, one looks for an ``~\emph{extremal}~'' of the action. Anyway, this distinction is not relevant
in what follows, and we shall neglect this kind of mathematical details.}   If the system had followed any 
different , non ``~correct~'', path, the value of the integral would have been 
larger.   \\
Let us briefly review the implications of the Principle in Physics. The most important consequence of the Principle is 
a set of equations, known as the Euler-Lagrange equations, describing as the system, encoded by $\mathcal{L}$, changes with 
time:                                                           
                                    
\begin{center}
\emph{Euler-Lagrange equations}$\longrightarrow$ \emph{time evolution of the system }
\end{center}

Once the Lagrange function $\mathcal{L}$ is assigned, the Euler-Lagrange equations allows to follow instant after instant its
time changes.  
A priori, different expressions of the Lagrangian might be conceived. The select the most appropriate Lagrangian for a definite 
physical situation one follows certain ``~recipes~'',  which basically consist in the requirements:\\

\begin{enumerate}
 \item mathematical \emph{simplicity};
 \item mathematical \emph{elegance}.
\end{enumerate}

The motivation for simplicity follows from the need for a physicist to extract from the theory experimentally testable 
predictions. From this point of view, if the model is mathematically so complex to make impossible to recover an
even approximate solution of the evolution equations, it is formally sophisticated but... useless!\\
The second request reflects the beliefe that \textit{``~Nature is beautiful~''} \cite{green} and often translates into the existence of
symmetry properties which strongly reduce the possible form of the Lagrangian.
For example, in Particle Physics it is mandatory to endow  the ``~simplest~'' possible Lagrangian with, at least, two
fundamental symmetries: 
\begin{itemize}
 \item a) Lorentz invariance, taking into account the fundamental laws of Special Relativity \cite{einstein}. The evolution equations must
maintain the same form when space and time are ``~mixed~'' together as Einstein proposed at the beginning of the $20^{th}$ century.
\\
\item b) Local Gauge Invariance, unifying the Fundamental Interactions among elementary particles in the mathematical framework
of Group Theory.
This principle, originally introduced by Fock and London \cite{supercond} to  describe the phenomenon of super-conductivity, 
leads to the general  rule  of  ``~minimal substitution~'' to couple every elementary particle to all known fundamental forces.
The term remarks, once again, how in physics beauty and simplicity are often synonymous.
 \end{itemize}

The immediate question is now: is there any experimental evidence of the assumed Principles?
 \\
The most spectacularly positive  answer to this question is undoubtedly provided by the recent discovery of the 
(or, maybe, one of the..) Higgs Boson at LHC. As it is known, the existence and the properties of this entity were postulated 
by Higgs \cite{higgs} as a necessary condition for the validity of the theoretical description that assumed the local gauge invariance of a 
special relativistically invariant Standard Model, essentially  based on the Minimal Action Principle. 
 In the formalism of Relativistic Quantum Field theory, the knowledge of the  Lagrangian allows to derive in a relatively 
simple way the expression of the cross section of a process where an initial state of two particles is transformed by the present 
Forces into a final two- particle state possibly different from the initial one. In practice, this leads to the prediction concerning 
the number of final states of a certain type observed after having performed a very great number of collisions between the initial 
particles.. These predictions have been successfully tested, first at the CERN LEP electron-positron collider  and afterwards at the 
Chicago Tevatron proton-antiproton and at the CERN LHC proton-proton colliders. In fact, apart from the production  of  final 
two particle states, the main hope of  these experiments was that of producing as intermediate state the fundamental Higgs boson. 
This discovery has been announced at LHC on July 2012, and its relevance has been stressed by the whole scientific community, a part 
of which (including the authors of this short article) considers the Higgs production as the scientific event of the $21^{th}$ Century.   

\section{ ``~Elevation~'' and ``~Awareness~''. }

It seems to us to be reasonably allowed at this stage to claim that the Principle of Minimal Action appears fairly confirmed by the 
most severe and accurate available experimental measurements of High Energy Physics. This means that the evolution of an elementary
matter state in the presence of the known Forces appears to follow those rules that are dictated by the initial acceptance of the 
Principle. In the following part of this paper we shall present our very personal view of a possible interpretation of this 
connection between the Principle and the matter system evolution, that would be strictly connected with the role that Goethe 
proposed to his Action.\\

The starting point of our discussion is the search of an intrinsic meaning to be associated with the two relevant energies, 
$T=$ Kinetic and $U=$ Potential, that a material system can have.  We shall first consider the simplest  case of a single elementary 
particle system (for instance, one electron).   Quite generally, we can say that the Kinetic Energy is associated to 
the motion of  the particle, thus a change of Kinetic Energy is only possible if the ``~velocity~'' of the particle changes 
 In particular, if the velocity increases, as a consequence of an acceleration, this energy increases as well, till a final 
value allowed by the reachable possible velocity (limited in the theoretical Einstein vision) of the particle.
 We like to interpret, with a touch of imagination,  an increase of the  Kinetic energy as a ``~desire~'' of  
the material (i.e. massive) particle of approaching its existing absolute unreachable  velocity limit 
(the velocity of the light). This process can be seen as a search for  \emph{Elevation}. In this very personal attitude, 
we shall give the Kinetic Energy the name of  \emph{``~Elevation Energy~''}.  \\

The nature of the Potential Energy is, in this view, different. This Energy  is only fixed by the position occupied at a certain 
time by the particle.  Its value would be different, for instance, if the particle were  located in different positions with the 
same velocity. These changes of $U$ would  not introduce therefore the previous kind of ``~Elevation~'' generated by changes of $T$. 
In the same  very speculative and personal approach that we are following,  we would give the Potential Energy $U$, 
that is only determined by the  particle \emph{``~awareness~''} of its position, the special name of  
\emph{``~Awareness Energy~''}.  \\
When moving to a less simple material system, generally made of several elementary particles, we shall maintain our view of the 
previous unconventional definitions of its two possible Kinetic and Potential Energies. This appears to us, at least qualitatively, 
an acceptable generalization of the proposal made in the elementary one particle case.\\


Accepting  the previous definitions, we can now derive an interpretation of the Minimal Action Principle. Rigorously speaking, 
the Principle tells that, in the ``~correct~'' motion of a material system from an initial time $t_1$ to a final time $t_2$, 
the difference between the total amount of Kinetic Energy and the total amount of Potential Energy   ``~accumulated~'' in the time 
interval is minimum. In our personal language, we would say:\\

 \emph{in the  ``~correct~'' motion of any material system in the presence of the known forces, the difference between the ``~accumulated~'' 
Elevation Energy $T$ and  the ``~accumulated~'' Awareness Energy  $U$ is minimum}.
\\

Until now, we are not saying anything new. I fact, we are only giving $T$ and $U$ , the real physical quantities of the particle, 
some arbitrary ``~nicknames~'', and re-expressing the Minimal Action Principle in the conventional way, in a rigorously 
physical language.
The next jump is produced by our personal  fancy, perhaps intuition, that leads us to replace $T$ and $U$ in the  previous 
formulation and write, in a new different  (Universal?) language:\\

\emph{in the ``~correct~'' motion of any material system in the presence of the known forces, the \emph{difference} between 
the ``~accumulated~''   elevation and the ``~accumulated~'' awareness}
\begin{center}                                    
 \emph{is minimum}.
\end{center}
The above statement represents our personal ``~Universal~'' reformulation of  Hamilton Principle of Minimal Action. 
Our vision of the Hamilton's Action principle stops now.\\
 The possibility of an interpretation that goes well beyond the 
physical treatment is, in our opinion, open and attractive. We have particularly in mind  the definition of Goethe's Action 
that we have found in a very recent article \cite{last}. This definition, that we quote here, finding  it definitely impressive, is:
\begin{quotation}
 ``~At the beginning there was the Action.  Goethe's Action represents the synthesis which is requested to grasp the essence of Faust, 
that in the poem must represent the image of the modern man. The Action is seen as the unique possible existential answer that mediates   
between the two most natural instincts of every man. These are the instinct of the challenge to God, that each man needs to follow his 
always bigger, unsatisfiable aspirations (~\emph{Elevation}~), and the moment of humility, in which the man must  remember his mortality 
and  return to his most genuine human values, to the relevance of  thinking for the community where he lives (~\emph{Awareness}~)~''
\end{quotation}

\section{Conclusions}
It is difficult for us not to find, from what exposed, an analogy between the roles of the Action in the Goethe poem and in the 
Hamilton Principle. The extra final point that we want to stress at the end of this extremely personal proposal is however the following 
one: in the physical formalism that we have summarized, the Action balances between the two alternative Energies in a very precise way, 
that is not present in the Goethe's definition. According to the Hamilton view that we have proposed, the ``~Accumulation~'' of Elevation 
and the ``~Accumulation~'' of Awareness cannot be separately arbitrary: otherwise stated, in the ``~correct~'' evolution, there cannot  
be in particular an increase of only one of the two extremes. Pure Elevation increase without Awareness increase is not 
``~Correct~'', and viceversa.


\begin{thebibliography}{99}
\bibitem{theogony} 
West \  M.\  L.\ \textit{Theogony}.\  Oxford University Press (1966), pp. 40
\bibitem{logos} G.\ Strecker, F.\ W.\ Horn, \textit{Theology of the New Testament}, (2000)\  pp.473 
\bibitem{faust} C.\ E.\ Passage
\textit{Goethe's Plays, by Johann Wolfgang von Goethe translation into English with Introductions},
 Benn Limited Pubbl.,\ (1980),\  ISBN 0510000878 / 9780510000875 / 0-510-00087-8
\bibitem{hamilton}
W.\ R.\ Hamilton, \textit{On a General Method in Dynamics},\ Phil.\ Transact.\ of the Roy.\  Soc.\  Part II (1834) pp.247 ;\ 
Part I (1835) pp.95
\bibitem{green}
B.\ Greene,\ \textit{The Elegant Universe: Superstrings, Hidden Dimensions and the Quest for the Ultimate Theory},\ 
Vintage;\  New Ed edition (3 Feb 2000)
\bibitem{einstein} A.\ Einstein,\  \textit{Zur Elektrodynamik bewegter K\`orper},\  Annalen der Physik,\ \textbf{17},\ (1905)\ pp. 891
\bibitem{supercond}
V.\ Fock,\ \textit{\`Uber die invariante Form der Wellen und der Bewegungsgleichungen f\`ur
einen geladenen Massenpunkt},\ Zeit.\ f\`ur\ Physik \textbf{39},(1926)\ pp.226\\
F.\ London,\ H.\ London, \textit{The Electromagnetic Equations of the Superconductor}. 
Proc.\ of the Roy.\ Soc.\ A:\ Math.,\ Phys.\ and\ Eng.\ Sciences,\ \textbf{149},\ (1935), \ pp. 866
\bibitem{higgs} P.\ W.\ Higgs  \textit{Broken Symmetries and the Masses of Gauge Bosons}. Phys.\ Rev.\ Lett. \textbf{13} (1964) pp. 508
\bibitem{last} M. Quadrini, \textit{Faust}, (2011), on-line article, http://mediacritica.it/2011/09/22/faust/
\end{thebibliography}
\end{document}